\documentclass[aip,apl,amsmath,amssymb,reprint,superscriptaddress]{revtex4-1}

\usepackage{graphicx}

\usepackage{xcolor}

\begin{document}

\title[]{Alignment of the Diamond Nitrogen Vacancy Center by Strain Engineering}

\author{Todd~Karin}
\affiliation{Department of Physics, University of Washington, Seattle, Washington 98195, USA}
\author{Scott~Dunham}
\affiliation{Department of Electrical Engineering, University of Washington, Seattle, Washington 98195, USA}
\author{Kai-Mei~Fu}
\affiliation{Department of Physics, University of Washington, Seattle, Washington 98195, USA}
\affiliation{Department of Electrical Engineering, University of Washington, Seattle, Washington 98195, USA}

\date{May 20, 2014}

\begin{abstract}
The nitrogen vacancy (NV) center in diamond is a sensitive probe of magnetic field and a promising qubit candidate for quantum information processing. The performance of many NV-based devices improves by aligning the NV(s) parallel to a single crystallographic direction. Using \emph{ab initio} theoretical techniques, we show that NV orientation can be controlled by high-temperature annealing in the presence of strain under currently accessible experimental conditions. We find that $(89\pm7)\%$ of NVs align along the [111] crystallographic direction under 2\% compressive biaxial strain (perpendicular to [111]) and an annealing temperature of 970$^\circ$C.
\end{abstract}

\maketitle


%

The prospect of nanoscale sensing at ambient conditions has spurred renewed interest in the mature field of point-defect physics.  Studying point-defect physics using \emph{ab initio} computational techniques expands our understanding of important crystal defects and widens the scope of applicability for defect based devices.
In this work, we show that the orientation of the nitrogen-vacancy (NV) center in diamond can be controlled by annealing in the presence of strain. 

\begin{figure}[bt]
\includegraphics{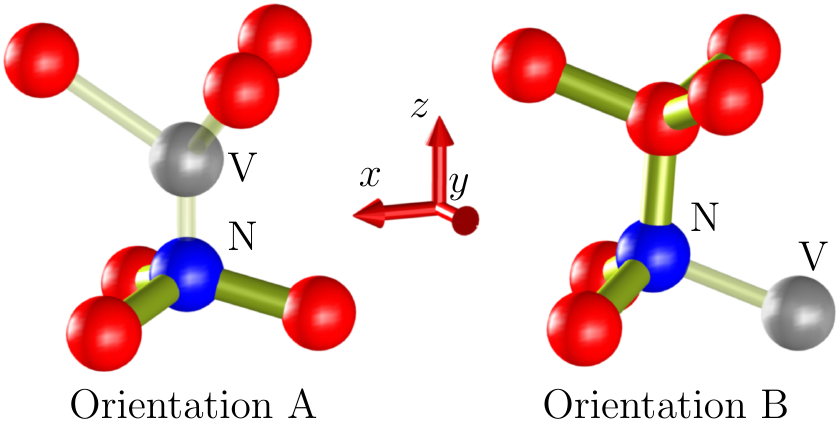}
\caption{Relaxed geometry of a nitrogen vacancy center with nearest neighbors shown for two NV orientations. Strain along the $z$ direction [111] causes the energies of the two orientations to differ. This energy difference can be exploited to polarize NV centers into a given orientation. The vacancy and bond-cylinders are given as guides for the eye. }
\label{fig:NVgeometry}
\end{figure}

The diamond NV center is a technologically-relevant and well-studied defect consisting of a substitutional nitrogen and nearest-neighbor vacancy (Fig.~\ref{fig:NVgeometry}). \cite{Schirhagl2014,Maze2011,Doherty2011} The center's electron spin-triplet ground state has an unusually long spin coherence time, exceeding 1 ms at room temperature.\cite{ref:balasubramanian2009usc} This long coherence time, coupled with the ability to perform optically detected magnetic resonance, enables the NV center to be used as a sensitive probe of temperature,\cite{ref:kuscko2013nst} electric field,\cite{ref:dolde2014nds} magnetic field,\cite{ref:balasubramanian2008nim, ref:maze2008nms, ref:taylor2008hsd, ref:grinolds2013nmi} strain,\cite{ref:grazioso2013mfs} and pressure.\cite{ref:doherty2014epm}  The NV center is also a promising qubit candidate for quantum information applications.\cite{ref:bernien2013heb, ref:maurer2012rtq, Waldherr2014}

Many NV-based devices would benefit from simultaneous control over the NV position and orientation in the host diamond.
Due to the tetrahedral coordination of the diamond lattice, the NV has four possible orientations, two of which are shown in Fig.~\ref{fig:NVgeometry}. Control over orientation increases NV homogeneity and improves the performance of sensors based on ensembles. In magnetic sensing for example, alignment inhomogeneity increases noise because only the magnetic field projection on the NV symmetry axis is measured.\cite{ref:taylor2008hsd,ref:degen2008smf}
Similarly, quantum information applications typically require qubits with identical properties, including alignment, to facilitate qubit coupling and entanglement generation.\cite{ref:cabrillo1999ces,ref:duan2001ldq,ref:benjamin2009pmb}
Finally, both sensing and quantum information may benefit from coupling the NV to an optical resonator, where NV dipole alignment to the resonant mode polarization is critical.\cite{ref:faraon2011rez,ref:faraon2012cnv, ref:hausmann2013cnv}

Current techniques to create NVs provide control over either NV position or orientation, but not both simultaneously. In chemical vapor deposition grown diamond, NV defects may align to a preferential direction during growth,\cite{ref:michl2014pap,Edmonds2012} however this method provides no control over spatial location. Conversely, implantation and annealing positions NVs to nanoscale accuracy,\cite{Toyli2010} but leaves them randomly oriented. Our method aligns pre-existing NV centers through strain engineering, thus enabling simultaneous control over NV position and alignment.

We demonstrate with \emph{ab initio} techniques that the NV orientation can be controlled by annealing in the presence of strain.
Strain breaks the symmetry between different NV orientations,
causing some orientations to be energetically preferred over others. By annealing the sample at a temperature where reorientation can occur, NVs preferentially align into lower energy directions. We show that the energetics of defect migration make this scheme possible, and quantitatively derive the degree of NV orientation expected for different strains and annealing temperatures.

There has been recent success using \emph{ab initio} techniques to confirm and predict the electronic structure of defects in diamond.\cite{Gali2013,Chanier2013,Hossain2008,Gali2009}  
Density functional theory (DFT) is particularly well-suited to studying the effect of strain on defect energies because energy differences between similar structures generally converge faster than absolute energies.\cite{Ahn2009,Ahn2006,Francis1990} 


To calculate the energy difference between NVs of various orientations under strain, simulations were performed using the \emph{ab initio} total-energy and molecular-dynamics program VASP (Vienna ab-initio simulation program) developed at the Instutut f\"ur Theoretische Physic of the Technische Universit\"at Wien. \cite{Kresse1994,Kresse1996, Kresse1996b}
NV defects were simulated using DFT in the generalized gradient approximation (GGA, PW91), \cite{Perdew1992,Perdew1993} chosen for its excellent prediction of stress effects in Si. \cite{Ahn2006,Ahn2009} The charge state of the NV was neutral throughout the calculation, as NV$^0$ is the predominant charge state formed by implantation and annealing prior to surface oxidation.\cite{Fu2010,Chu2014} To minimize defect-defect interactions between neighboring supercells, we used a large 647 atom supercell. Simulations continued until the total energy changed less than 0.01~eV per iteration.  The plane wave energy cut-off was 348~eV, as increasing to 400~eV modified the energy of a non-strained supercell by only 10~meV. 
To assess convergence in \(k\)-points for the strain calculation, we compared the energies of orientations A and B at zero strain, where the two orientations should have the same energy by symmetry. We used a 2x2x2 \(k\)-mesh, which had \( |E_B-E_A| = 1.8\)~meV (while 1x1x1 had an error of 16.4~meV).

We first used VASP to calculate the lattice constant for bulk diamond as a starting point for the defect simulations, obtaining \(a_o=3.560\)~\AA, close to the experimental value \(a_o=3.567\)~\AA.\cite{Holloway1991} Because of the perfect translational invariance in bulk diamond, a small 24 atom supercell and 10x10x10 Monkhorst-Pack $k$-mesh were used to find the lattice constant (with good convergence in $k$ points).

\begin{figure*}
\includegraphics{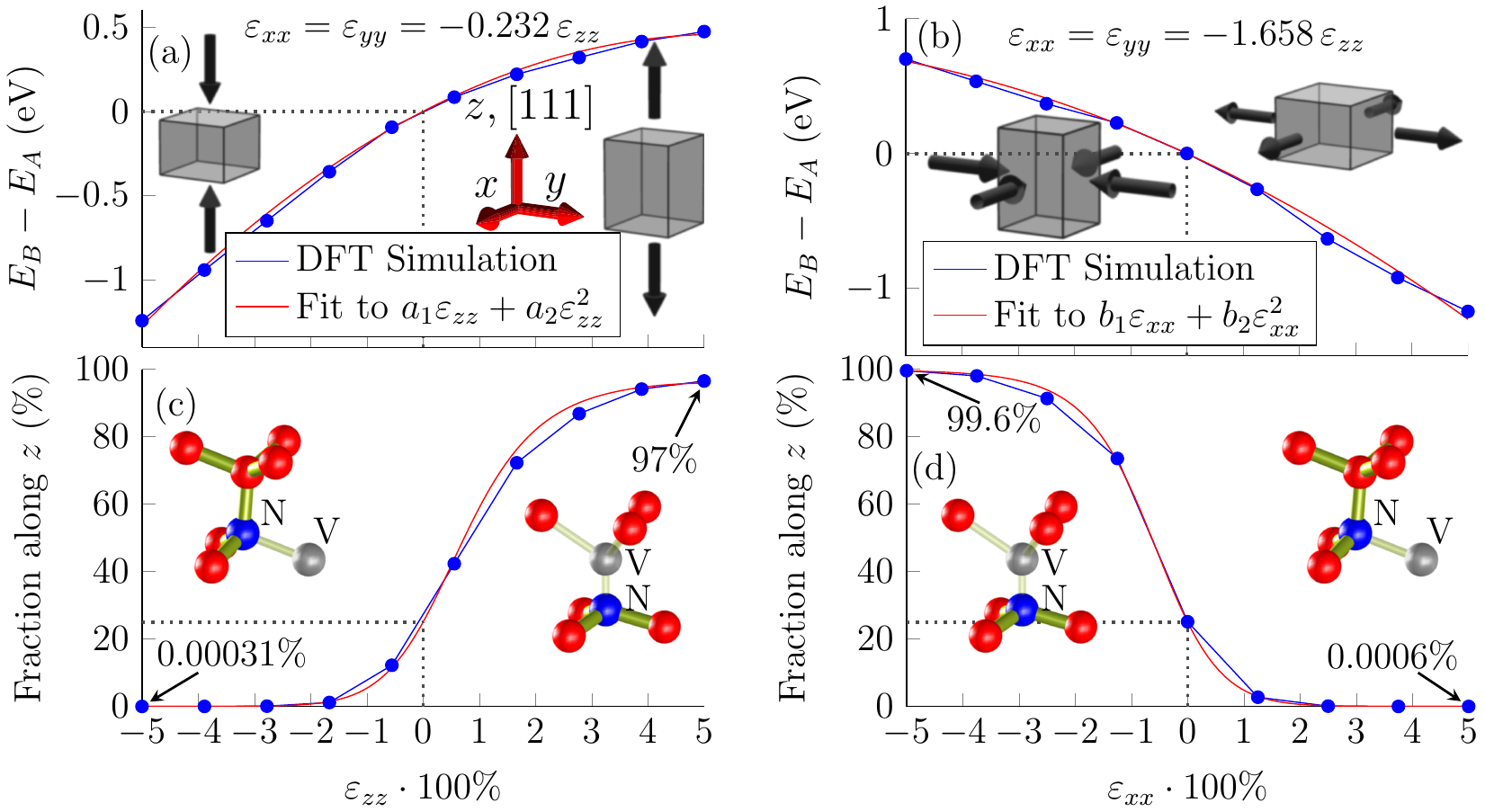}
\caption{
\textbf{(a)} Effect of uniaxial strain in $z$ and zero-stress condition in $x,y$ on the energy difference between orientations A and B as shown in Fig.~\ref{fig:NVgeometry}. Tensile strain along $z$ results in the $z$ oriented NV to be preferred. At zero strain, the energies of the two orientations should be equal by symmetry, as depicted by the dotted lines. A fit to a second order polynomial is provided to easily reproduce the calculated values. The best fit parameters are \(a_1 = (17.3 \pm 0.6)\) eV, \(a_2 = (-160 \pm 10)\) eV, with uncertainties due to the fit only. Insets depict the direction of strain applied by arrows.
\textbf{(b)} Same as a, except strain is applied biaxially in \(x,y\), with zero stress condition in $z$. Biaxial compressive strain results in the the $z$ oriented NV to be preferred. The fit line uses $b_1 = (-19 \pm 1)$ eV, $b_2 = (-110 \pm 30)$ eV.
\textbf{(c)} Assuming an annealing temperature of 970\(^\circ\)C, the predicted fraction of NVs oriented along \(z \propto [111]\) can be calculated. Because the energies of the two orientations are equal at zero strain, the fraction oriented along [111] should be 25\%, depicted by the dashed lines. The insets show the preferred NV orientation in each strain regime.
\textbf{(d)} Same as c, except for strain applied biaxially along \(x,y\).
}
\label{fig:strainSweep}
\end{figure*}

To determine how strain affects NV orientation, we found the difference in energy between two NV orientations for a range of strain fields (Fig.~\ref{fig:NVgeometry}).  (We do not distinguish between NVs with anti-parallel orientations.) The $x$,$y$ and $z$ axes of the simulation volume were oriented along \([1\bar{1}0]\), \([11\bar{2}]\) and \([111]\) respectively.
These directions were chosen based on the symmetry of the NV. Uniaxial strain along [111] makes the [111] orientated NV different from other orientations, thus possibly leading to an energy difference favoring the [111] orientation.
In the first simulation, $z$ uniaxial strain \(\varepsilon_{zz}\) was applied, with the $x,y$ strain given by the expected behavior of bulk diamond unconstrained in $x,y$ (Fig.~\ref{fig:strainSweep}a).
Using the experimentally determined elastic constants of bulk diamond, we find that under uniaxial strain in \(z\), bulk diamond acquires a strain of \(\varepsilon_{xx} = \varepsilon_{yy} = -0.232\,\varepsilon_{zz} \), with all other \(\varepsilon_{ij} = 0\).
\footnote{See supplemental material at [URL will be inserted by AIP] for derivation of linear strain response coefficients.}
In the second simulation (Fig.~\ref{fig:strainSweep}b), biaxial strain was applied in \(x,y\), with the $z$ dimension similarly given by the bulk diamond elastic constants (\(\varepsilon_{xx} = \varepsilon_{yy} = -1.658\, \varepsilon_{zz} \), all other \(\varepsilon_{ij} = 0\)). Because the strain splitting of NV orbitals is highly linear up to 60 GPa hydrostatic pressure ($\sim$4.5\% strain),\cite{Doherty2014} our use of linear elastic theory in the current work is well justified.

The simulation results for the energy difference between orientation A and B as a function of strain are shown in Figs.~\ref{fig:strainSweep}a,b. The relative smoothness of the energy vs. strain curve implies that Pulay stresses (a numerical artifact) are negligible.\cite{Francis1990} The degree of NV orientation depends on the energy difference between the two orientations relative to the annealing temperature where NV realignment can occur. Because NV realignment is a precursor to NV diffusion, we estimated the effective energy barrier for NV realignment using the temperature at which NV luminescence disappears, 1300-1500$^\circ$C. \cite{Pezzanga2011,Pinto2012} At these temperatures, NVs diffuse to form defect clusters. Since NV diffusion to clusters requires many defect reorientations and migrations, we will use $(1300 \pm 100)^\circ$C as an estimate for the temperature at which NV reorientation happens at a rate of 1~s$^{-1}$. The rate at which defect reorientations occur is
\begin{equation}\label{eq:rate} R = \nu e^{-W/k_B T}, \end{equation}
where \(\nu\) is the attempt frequency of reorientation and \(W\) is the effective energy barrier for reorientation.\cite{Pinto2012} The attempt frequency \(\nu\) can be estimated by dimensional analysis given the known attempt frequency for vacancy migration in silicon.\cite{El-Mellouhi2004} Due to the increased stiffness of diamond over silicon, the attempt frequency in diamond (\(3 \times 10^{13}\,\text{s}^{-1}\)) is around 3 times higher than that of silicon. Using these values, the effective energy barrier for reorientation of an NV is $(4.2 \pm 0.3)$ eV, slightly lower than the value of 4.85 eV found using DFT.\cite{Pinto2012}.

Annealing at a temperature \(T\) causes the NV population to exponentially approach its thermal equilibrium orientation at the rate \(R\). Since only a few reorientations are required for the NV population to reach thermal equilibrium with respect to orientation, we used a slow reorientation rate \(R=\text{hour}^{-1}\) to calculate the proper annealing temperature for NV alignment, $970^\circ$C. Reorientation will necessarily be accompanied by diffusion because the energy barriers for reorientation and diffusion are similar.\cite{Pinto2012} However, if the annealing time is limited to a few reorientation time constants, the amount of diffusion will be less than a few lattice sites. We assume that the energy barrier is roughly independent of temperature, motivated by the temperature-independent barrier for vacancy diffusion in silicon.\cite{Watkins2008} The energy barrier may depend on strain, which we account for using an uncertainty in the energy barrier of \(\pm 1\)~eV at 2\% strain.\cite{Pinto2012}

The fraction of NVs aligned parallel to $z\propto [111]$ can be calculated from the energy difference between NV orientations and the annealing temperature. Noting the triple degeneracy of other orientations, the fraction of NVs aligned parallel to $z$ is
\begin{equation}\label{eq:Nz} N_z = \frac{1}{3 e^{- (E_B-E_A)/k_B T} + 1 } \end{equation}
where \(k_B\) is the Boltzmann constant and \(T\) is the annealing temperature. The use of energy instead of the free energy in Eq.~\ref{eq:Nz} assumes that the entropy of each NV orientation is the same. This is likely a good approximation even under strain given that the arrangement of atoms is highly similar for the different orientations. However, we note defect entropies could be calculated from the phonon density of states.\cite{Lucas2009} 

Figs.~\ref{fig:strainSweep}c,d show the fraction of NVs aligned parallel to $z$ (Eq.~\ref{eq:Nz}) vs. strain under annealing at 970$^\circ$C for the two strain directions studied. We find that either uniaxial expansion in \(z\) or biaxial compression in \(x,y\) favor the $z$ orientation of NV (Fig.~\ref{fig:strainSweep}c,d). Impressively, at 5\% compressive biaxial strain, $(99.5^{+0.5}_{-4.7})$\% of NVs align along~$z$. At 2\% compressive biaxial strain, $(89 \pm 7)$\% of NVs align along $z$. The uncertainty here is estimated from the uncertainty in the strain dependence of the reorientation barrier, which we found to create a larger uncertainty than the temperature where NV luminescence disappears or choice of DFT functional. In order to test whether the choice of functional affects the result, we checked several strain calculations using the local density approximation (LDA).\cite{Perdew1981} We found that the LDA result differed from the GGA one by on average 12~meV.

The effect of strain fields on the energy of an NV orientation can be phenomenologically understood by studying the optimal relaxation of a supercell with an NV. We relaxed the size and shape of a 647 atom supercell containing a \(z\)-oriented NV using an increased plane wave cutoff of 426~eV. Compared to a 648 atom supercell of bulk diamond, the NV supercell relaxes to a strain of
\begin{equation*}
\varepsilon = \left( \begin{array}{c c c}
-2.1     &    0  &       0  \\
      0  &  -1.8  & 0.4 \\
      0  & 0.4  &  3.7
  \end{array}
  \right) \cdot 10^{-4}.
\end{equation*}
The small off diagonal strains are attributed to the asymmetry of the supercell.
These results imply that NV energy lowers with tensile strain along the NV axis and biaxial compressive strain perpendicular to the axis, and can be used to predict the NV response to arbitrary strains.

Several experimental techniques can create biaxial or uniaxial strain in diamond. Biaxial strain can be created by selectively damaging diamond with an ion beam in an annulus surrounding an NV. The damaged material swells compared with the undamaged diamond, thus applying biaxial compressive strain to the NV.
In recent experiments, a strip of damage created with an ion-beam created up to $\sim$10 GPa of tensile stress ($\sim$1.5\% strain) near the damaged region.\cite{Olivero2013} Uniaxial compressive strain can be achieved in a diamond anvil cell, resulting in the depletion of one NV orientation. Since diamond anvil cells are capable of very high pressures (hundreds of GPa), this technique is likely limited by the yield stress of diamond. In real-world materials, non-hydrostatic stresses can fracture materials by activating slip systems, therefore experimentally achievable strains for both techniques may be limited to \(\sim\)2\%.

In conclusion, we have shown that the orientation of the diamond NV center can be manipulated by annealing in the presence of strain at realistic temperatures and strain fields. This technique allows NVs to be created at precise locations and subsequently aligned in desired directions. This work presents an additional dimension of control for engineering NV-based devices, and advances the emerging fields of defect-based quantum information processing and nanoscale sensing.

This work was supported by the National Science Foundation under Grant No. 1342902, DGE-0718124 and DGE-1256082. This work used the Extreme Science and Engineering Discovery Environment (XSEDE), which is supported by National Science Foundation grant number ACI-1053575.

\bibliography{toddkarin.bib,fu_lab_bib.bib}

\end{document}


\title[]{Alignment of the Diamond Nitrogen Vacancy Center by Strain Engineering:\\ Supplemental Material}

\author{Todd~Karin}
\affiliation{Department of Physics, University of Washington, Seattle, Washington 98195, USA}
\author{Scott~Dunham}
\affiliation{Department of Electrical Engineering, University of Washington, Seattle, Washington 98195, USA}
\author{Kai-Mei~Fu}
\affiliation{Department of Physics, University of Washington, Seattle, Washington 98195, USA}
\affiliation{Department of Electrical Engineering, University of Washington, Seattle, Washington 98195, USA}



\date{May 20, 2014}

\begin{abstract}
\end{abstract}

\maketitle

The response of a bulk diamond crystal to strain along $z\propto [111]$ can be found by transforming the rank-four elastic constants tensor in the original coordinate system ([100],[010],[001]), 
\[ 
c = \left(\begin{array}{cccccc}
c_{11} & c_{12} & c_{12} & 0 & 0&0 \\
c_{12} & c_{11} & c_{12} & 0 & 0&0 \\
c_{12} & c_{12} & c_{11} & 0 & 0&0 \\
0&0&0&c_{44}&0&0 \\
0&0&0&0&c_{44}&0 \\
0&0&0&0&0&c_{44} \\
 \end{array}
 \right),
\]
%
into the new coordinate system (\([1\bar{1}0],\,[11\bar{2}],\,[111]\)).\cite{Nye} Here we have written the strain tensor in its reduced form, where components 1,2,3 correspond to uniaxial strain along x,y,z and 4,5,6 to shear strains along $yz, xz, xy$. \cite{Dresselhaus} 
The generalized Hooke's law strain energy is \[W = \sum_{ij}\frac{1}{2} \varepsilon_i c_{ij} \varepsilon_j .\] If a strain \(\varepsilon_3\) is applied in the $z$ direction, the diamond will accumulate a strain in $x,y$ in order to minimize the strain energy:
\[ \varepsilon_1 = \varepsilon_2 = -  \overbrace{\frac{c_{11} +2 c_{12} -c_{44}}{2c_{11}+4c_{12}+c_{44}} }^{0.232 \pm 0.002} \varepsilon_3 , \quad \varepsilon_4 =\varepsilon_5 = \varepsilon_6 = 0.\]
Here we have used experimentally determined elastic constants for diamond.\cite{McSkimin1972b}
Alternately, if biaxial compressive strain \(\varepsilon_1 = \varepsilon_2\) is applied, then the diamond acquires a strain 
\[ \varepsilon_3  = - \overbrace{\frac{c_{11} +2 c_{12} -c_{44}}{c_{11}+2c_{12}+2c_{44}} }^{0.302 \pm 0.003} (\varepsilon_1+ \varepsilon_2), \quad \varepsilon_4 =\varepsilon_5 = \varepsilon_6 = 0. \] These proportionality constants were used to deform the supercell given an applied uniaxial \(z\) or biaxial \(x,y\) strain.

\bibliography{toddkarin.bib,fu_lab_bib.bib}